\documentclass[%
reprint,
superscriptaddress,
amsmath,amssymb,
aps,
]{revtex4-2}

\usepackage{graphicx}
\usepackage{dcolumn}
\usepackage{bm}
\usepackage{verbatim}   
\usepackage{physics}
\usepackage{amsmath}
\usepackage{xcolor}
\usepackage{float}

\begin{document}

\preprint{APS/123-QED}

\title{A Source of Deterministic Entanglement for Matter-Wave Networks}

\author{Chen Li}
\email{chen.li@tuwien.ac.at}
\affiliation{Vienna Center for Quantum Science and Technology, Atominstitut, TU Wien, Stadionallee 2, 1020 Vienna, Austria}

\author{RuGway Wu}
\affiliation{Vienna Center for Quantum Science and Technology, Atominstitut, TU Wien, Stadionallee 2, 1020 Vienna, Austria}

\author{J\"{o}rg Schmiedmayer}
\affiliation{Vienna Center for Quantum Science and Technology, Atominstitut, TU Wien, Stadionallee 2, 1020 Vienna, Austria}

\date{\today}

\begin{abstract}

We describe a deterministic and experimentally feasible protocol for generating entangled pairs of ultracold neutral atoms through controlled dissociation of diatomic Feshbach molecules. The dissociation process naturally produces nonlocal quantum correlations in spin, position-momentum, and path degrees of freedom, enabling the deterministic preparation of Einstein-Podolsky-Rosen pairs of massive particles and multiqubit states through hyperentangled encoding. Having each atom of the pair prepared in a matter waveguide, the scheme can be scaled to hundreds of parallel entanglement sources in an array connected to a matter wave optical network of beam splitters, phase shifters, interferometers, tunnel junctions and local detectors. The protocol builds on established techniques, including programmable optical potentials, high-fidelity single-particle control, single-molecule initialization, controlled molecular dissociation, and quantum gas microscopy with near-perfect detection, making it directly implementable with current technology. The proposed architecture naturally integrates with atomtronics circuits and chip-based matter-wave optics, offering a deterministic entanglement source for quantum nonlocality tests, precision metrology, and scalable neutral-atom quantum processors.

\end{abstract}

\maketitle

\section{Introduction}
\label{sec:intro}

The generation, manipulation, and detection of entanglement are fundamental to quantum technologies, including quantum simulation, quantum communication, quantum computing, and quantum metrology~\cite{Reid2009, Acin2018}. Photonic systems, in particular, have played a pioneering role: entangled photon pairs produced via spontaneous parametric down-conversion (SPDC) have enabled landmark demonstrations of the Einstein–Podolsky–Rosen (EPR) paradox and Bell inequality violations~\cite{Pan2012, Rudolph2017}. However, the probabilistic nature of SPDC sources and the limited efficiency of photon detectors remain major challenges for scaling entanglement to large systems and networks with many nodes, motivating the exploration of alternative platforms that combine deterministic entanglement generation with high-fidelity detection. 

Ultracold atoms and molecules provide precisely such an opportunity. Neutral atoms can be trapped, transported, and coherently manipulated with long coherence times and nearly ideal isolation from the environment~\cite{Henriet2020,Bluvstein2022}. Interactions between atoms can be precisely tuned via Feshbach resonances~\cite{Chin2010}. Deterministic state preparation and entangling operations are achieved by combining this tunability with advanced trapping and manipulation techniques. Crucially, quantum gas microscopy achieves single-atom resolution with detection fidelities above 99\%~\cite{Bucker2009, Bakr2009, Sherson2010, Omran2015, Bergschneider2018, Xiang2025, Yao2025, Jongh2025}. 
Significant progress has already been made in realizing atomic qubits and entanglement. Experiments have demonstrated strong squeezing and nonclassical correlations in both internal and external degrees of freedom of ultracold atoms~\cite{Esteve2008, Riedel2010, Maussang2010, Bucker2011, Peise2015, Schmied2016, Lange2018, Fadel2018, Borselli2021}. High-fidelity single-qubit gate operations have been demonstrated in optical arrays~\cite{Weitenberg2011,Xia2015,Wang2015,Wang2016}. Two-qubit interactions have been realized using collisional~\cite{Mandel2003,Anderlini2007,Kaufman2015,Yang2020,Zhang2023,Bojovic2025,Zhu2025} and Rydberg blockade mechanisms~\cite{Jaksch2000,Urban2009,Saffman2010,Levine2019}. 
Together, these capabilities establish ultracold atoms as a versatile platform for quantum metrology, simulation, and quantum computation. 

Previous proposals suggested using molecular dissociation as a source of entangled atom pairs~\cite{Fry1995, Opatrny2001, Kheruntsyan2005, Kheruntsyan2006, Zhao2007, Gneiting2008, Gneiting2010, Esquivel2011, Eckart2023}. In this paper, we move beyond conceptual proposals and present a fully detailed and experimentally feasible protocol, directly implementable with current cold-atom techniques. Our scheme not only enables the deterministic generation of entangled atom pairs with single-atom addressing and detection, but also opens a path towards extending to hundreds of parallel waveguides, offering a clear path to scalability. We provide a comprehensive analysis of correlation measurements in spin, position–momentum, and path, as well as hyperentanglement across multiple degrees of freedom. Although the gate speeds and fidelities are lower than those achieved in typical Rydberg quantum computing platforms~\cite{Levine2018, Levine2019, Bluvstein2022, Chiu2025, Bluvstein2025}, our approach offers distinct advantages: it deterministically generates Bell states of massive particles in several entangled degrees of freedom simultaneously. This establishes a powerful complementary platform to photonic systems, with strong potential both for exploring fundamental quantum correlations and for developing scalable neutral-atom architectures for quantum information processing.

\section{Experimental overview}
\label{sec:exp}

\begin{figure}[!htb]
\center
\includegraphics[width=0.75\columnwidth]{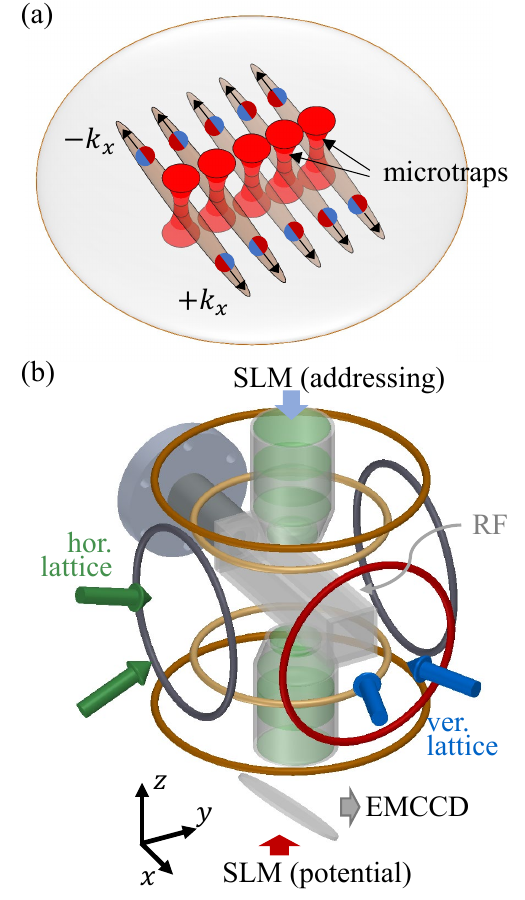}
\caption{(a) Principle of the protocol: single Feshbach molecules prepared at the centers of one-dimensional waveguides are dissociated into two atoms that propagate in opposite directions in entangled states. (b) Example implementation on a typical $^6$Li quantum gas platform: a two-dimensional optical lattice defines the waveguides, while two independent SLMs provide programmable potentials and local addressing. High-NA objectives are used both to project the SLM patterns and to perform single-atom–resolved fluorescence imaging. 
}
\label{fig:setup}
\end{figure}

The proposed protocol builds on established techniques in ultracold atom experiments, including Feshbach association and dissociation of molecules, programmable optical potentials using spatial light modulators (SLMs)~\cite{Nogrette2014, Chew2024} and digital micromirror devices (DMDs)\cite{Zupancic2016, Gauthier2016, Stuart2018, Tajik2019}, quantum gas microscopy with near-unity detection fidelity, and precise control of atomic ensembles down to the single‐particle level. These capabilities make the protocol readily implementable. 

\subsection{Experimental scheme}

Fig.~\ref{fig:setup}a illustrates the key elements of the protocol. Single diatomic molecules are first prepared and cooled in microtraps located at the centers of waveguides, which are arranged parallelly in the focal plane of a high-resolution imaging system. Upon triggered dissociation, the two atoms of each pair are launched into opposite directions along the waveguide, forming an entangled state in spin, spatial and motional degrees of freedom. By tailoring the waveguide geometry—such as merging, crossing, or branching—one can implement beam-splitter–style operations and controlled interactions between adjacent tubes. These operations allow the engineering and probing of entanglement. 

For concreteness, we illustrate the protocol using an example experimental setup shown in Fig.~\ref{fig:setup}b, based on a typical Lithium-6 quantum gas platform. $^6$Li atoms are particularly well suited for our protocol for several reasons. Logical qubits $\mid\uparrow\rangle$ and $\mid\downarrow\rangle$ can be encoded in two of the three ground-state sublevels of the $m_s = -1/2$ Paschen–Back manifold (labeled $\vert 1\rangle$, $\vert 2\rangle$, and $\vert 3\rangle$). These states can be coherently coupled via radio-frequency (RF) transitions. 
Broad Feshbach resonances in $^6$Li enable fine-tuning of interatomic interactions and the creation of tightly bound molecules~\cite{Strecker2003, Zurn2013Precise}, whose stability is enhanced by the suppression of inelastic collision losses~\cite{Petrov2004,Petrov2005}. At high magnetic fields, each spin state features nearly closed optical transitions for imaging, and the sensitivity of transitions between the two qubit states to magnetic field variations and noise is suppressed by almost three orders of magnitude compared to electron spins, greatly enhancing both detection fidelity and coherence time. 

In this example setup (Fig.~\ref{fig:setup}b), the waveguides are formed by two-dimensional optical lattices. Atoms in two balanced spin states are confined in a single layer of the vertical lattice (green arrows). A horizontal lattice (blue arrows) structures the trap into an array of one-dimensional waveguides with tunable spacing, balancing scalability and imaging resolution. Tightly focused microtraps are created using a SLM (potential), sharing the same high-numerical-aperture objective that also provides single-atom–resolved imaging. The SLM (potential) can further imprint complex atom-optics circuits, such as tunable waveguide links, or it can even define the entire waveguide network without the need for a horizontal lattice. On the opposite side, a separate objective combined with a SLM (addressing) enables local addressing and control of individual atoms. Magnetic coils generate homogeneous bias fields for tuning interactions, as well as field gradients for applying external potentials.

\subsection{Initialization of single Feshbach molecule}

\begin{figure}[!tb]
\center
\includegraphics[width=0.9\columnwidth]{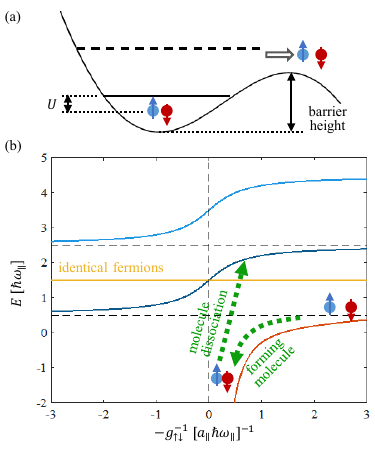}
\caption{(a) Preparation of a single fermion pair in a microtrap via the spilling technique: a tilted potential leaves only the lowest state trapped. $U$ denotes the mean interaction energy per particle. (b) Relative-motion spectrum of two atoms in a pair~\cite{Zurn2012}. Molecular states lie below the free-atom level, with binding energies set by the coupling constant $g_{\uparrow\downarrow}$. Green dashed arrows indicate molecular association and dissociation. 
}
\label{fig:preparation}
\end{figure}

Fermionic atoms provide an ideal platform for on-demand, scalable qubit initialization, as the Pauli exclusion principle inherently guarantees high fidelity~\cite{Raizen2009}. In tightly focused microtraps, the large vibrational energy gap allows precise control over the number of trapped atoms. Using a ‘spilling’ procedure (Fig.~\ref{fig:preparation}a), a calibrated magnetic-field gradient deforms the optical potential so that only a single fermion pair remains in the ground state. Atoms occupying higher‐energy levels spill out of the trap, after which the original potential is restored~\cite{Serwane2011}. This technique has achieved 97\% fidelity in preparing single-atom pairs per microtrap~\cite{Bayha2020}.

Fig.\ref{fig:preparation}b displays the kinetic energy of the relative motion of two atoms in the $\mid\uparrow\rangle$ and $\mid\downarrow\rangle$ states~\cite{Zurn2012}. Once each microtrap has been prepared with an atom pair, we sweep the magnetic field across the resonance to adiabatically transfer the pair into the ground state, forming a bound dimer with binding energy set by the 1D coupling constant $g_{\uparrow\downarrow}$~\cite{Zurn2013Pairing}. This controlled association initializes every microtrap with a single Feshbach molecule, which then serves as the entanglement source for the subsequent dissociation and EPR‐pair generation. 

\subsection{Molecular dissociation}

\begin{table*}[!htp]
\centering
\caption{Comparison of dissociation methods for $^6$Li Feshbach molecules. }
\label{tab:dissociation-methods}
\begin{tabular}{lcccc}
\hline\hline
Method &Timescale & Key features & Limitations \\
\hline
RF spin-flip  &  sub-ms & Fast, coherent, minimal recoil & Requires strong, homogeneous RF fields \\
Magnetic-field sweep & $0.5$–$10$ ms & Deterministic, tunable momentum spectrum & Relatively slow \\
Optical photodissociation & ns–$\mu$s & Fast, precise timing & Photon recoil; spontaneous emission \\
\hline\hline
\end{tabular}
\end{table*}

The dissociation of a single Feshbach molecule into two free atoms can be achieved by several alternative techniques, each with distinct advantages and limitations. Applying an RF $\pi$ pulse to transfer one spin state to another spin level quenches the interatomic interaction, thereby projecting the molecule into an unbound scattering channel~\cite{Regal2003}. This method is particularly suitable for $^6$Li; for example, an RF pulse couples the $\vert 1 \rangle$–$\vert 2 \rangle$ bound state to the $\vert 1 \rangle$–$\vert 3 \rangle$ continuum and introduces finite momenta to the atoms (Fig.\ref{fig:preparation}b). Note that the qubit encoding is redefined from $\vert 1 \rangle$–$\vert 2 \rangle$ ($\mid\uparrow\rangle$–$\mid\downarrow\rangle$) to $\vert 1 \rangle$–$\vert 3 \rangle$ ($\mid\uparrow\rangle$–$\mid\downarrow\rangle$). The process takes place on sub-ms timescales. Alternatively, a magnetic-field sweep across the dissociation threshold adiabatically converts the bound dimer into free atoms and allows precise control of the relative momentum distribution through the sweep rate~\cite{Mukaiyama2004,Durr2004,Greiner2005,Zhao2007}, operating on a relatively slower timescale of $0.1$–$10$ ms. A third approach, optical photodissociation, is implemented by driving the molecular state into the continuum with a narrow-band laser pulse~\cite{McDonald2016}. This technique offers ultrafast (ns–$\mu$s) and temporally precise triggering but unavoidably introduces photon recoil and a non-negligible probability of spontaneous emission, rendering it unsuitable for the present application. In all cases, the choice of method must balance speed, fidelity, and technical feasibility, as summarized in Table~\ref{tab:dissociation-methods}.

The dissociation process yields a maximally entangled spin singlet state of the two atoms, $(\mid\uparrow\downarrow\rangle - \mid\downarrow\uparrow\rangle)/\sqrt{2}$. Once the microtrap is switched off, the two atoms propagate along the waveguide with opposite, correlated momenta determined by the relative motion wavefunction. As they separate, their spatial wavepackets become distinguishable, allowing individual addressing in both manipulation and detection. Labeling the two spatial modes as `L' and `R', the state can be expressed as $(\mid\uparrow\rangle_L \mid\downarrow\rangle_R - \mid\downarrow\rangle_L \mid\uparrow\rangle_R)/\sqrt{2}$. This spatial separation thus converts the spin singlet into a nonlocal Bell pair of two atoms in separate waveguide channels. 

Molecular dissociation produces atom pairs that are not only entangled in spin, but also in position–momentum and path modes, and can even be hyperentangled across multiple degrees of freedom. In Sec.~\ref{sec:Measurements}, we demonstrate how these entanglements are measured. 

\subsection{Single-qubit operations on spin states}

\begin{figure*}[!tb]
\center
\includegraphics[width=2\columnwidth]{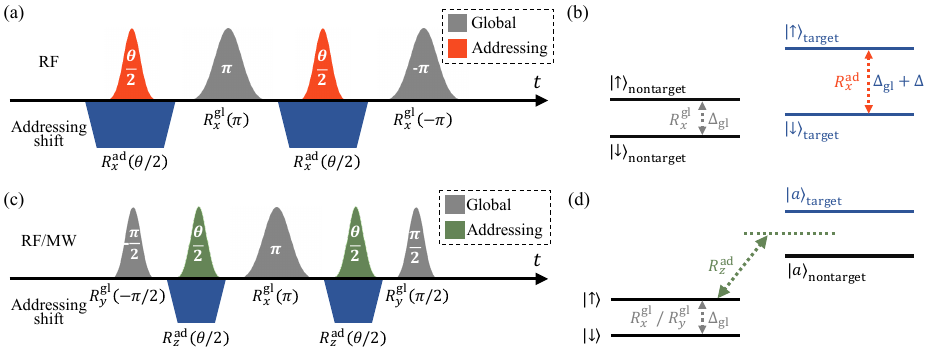}
\caption{Single-qubit gates in two schemes. (a, b) Scheme I: Single-qubit rotations are implemented by applying local AC Stark shifts to target atoms, with RF pulses resonant to the shifted transitions. (c, d) Scheme II: Single-qubit rotations are realized by coupling the qubit state $\mid \uparrow \rangle$ to an auxiliary state $\vert a\rangle$. An addressing laser shifts $\vert a\rangle$, so that off-resonant RF/MW pulses induce differential AC Zeeman shifts between target and non-target atoms, enabling addressed operations while keeping the system within the qubit subspace.
}
\label{fig:gate}
\end{figure*}

Spin rotations applied to the entire ensemble are defined as global operations, while those targeting only selected subsets are referred to as addressed operations. These addressed operations are realized via tightly focused laser beams—shaped by programmable optical modulators—that induce local AC Stark shifts on target atoms. When combined with a globally applied RF field, this technique allows for selective and coherent manipulation of individual qubits without perturbing the rest of the system~\cite{Weiss2004}. Such addressed control has been extensively demonstrated in neutral-atom platforms~\cite{Schrader2004, Weitenberg2011, Xia2015, Wang2015, Wang2016, Lee2013, Piltz2014}.

Fig.~\ref{fig:gate} presents two schemes for implementing an addressed $R_x (\theta)$ gate operation, which rotates a target quantum state by an angle $\theta$ about the $x$-axis of the Bloch sphere (with the $x$–$y$ plane defining the equatorial plane).

Fig.~\ref{fig:gate}a illustrates Scheme I. The addressed gate operation $R_x^\mathrm{ad} (\theta)$ is realized by combining two addressed half-rotations $R_x^\mathrm{ad} (\theta/2)$ with global spin-echo rotations $R_x^\mathrm{gl}(\pi)$ and $R_x^\mathrm{gl}(-\pi)$:
\begin{equation} 
R_x^{\mathrm{ad}} (\theta) = R_x^{\mathrm{ad}} \left(\frac{\theta}{2}\right) \cdot R_x^{\mathrm{gl}} (\pi) \cdot R_x^{\mathrm{ad}} \left(\frac{\theta}{2}\right) \cdot R_x^{\mathrm{gl}} (-\pi)\,. 
\label{eq:gate1} 
\end{equation}

The blocks below the time axis represent the energy shifts induced by the addressing laser beams, which are linearly ramped up to the target value, maintained during the addressing RF pulses, and then ramped down. The addressing RF pulses (orange, above the time axis) are resonant with the qubit transition shifted by the addressing laser beams (see Fig.~\ref{fig:gate}b), whereas the global RF pulses (gray) are resonant with all atoms when the addressing laser beams are off.
The addressing RF pulses selectively manipulate the target atoms while leaving the quantum superposition of non-target atoms unperturbed. This is ensured by the large differential shift $\Delta$ induced by the addressing laser — typically 10 kHz, an order of magnitude larger than the half-width at half-maximum (HWHM) of the transition, which is dominated by the Fourier width of RF pulses. Off-resonant AC Zeeman phase shifts induced by the addressing beams are compensated by implementing spin-echo sequences.  
Throughout the gate operation, non-target atoms experience time-reversed phase shifts, leading to negligible crosstalk.

Fig.~\ref{fig:gate}c illustrates Scheme II for the $R_x (\theta)$ gate. The rotation about the $x$-axis is realized by combining two addressed $R_z^{\mathrm{ad}} (\theta/2)$ gate with global $R_x^\mathrm{gl}$ and $R_y^\mathrm{gl}$ rotations (including the spin-echo sequence):
\begin{equation} 
R_x^{\mathrm{ad}} (\theta) = R_y^{\mathrm{gl}} \left(-\frac{\pi}{2}\right) \cdot R_z^{\mathrm{ad}} \left(\frac{\theta}{2}\right) \cdot R_x^{\mathrm{gl}} (\pi) \cdot R_z^{\mathrm{ad}} \left(\frac{\theta}{2}\right) \cdot R_y^{\mathrm{gl}} \left(\frac{\pi}{2}\right)\,. 
\label{eq:gate2} 
\end{equation}
As shown in Fig.~\ref{fig:gate}d, $R_x^\mathrm{gl}$ and $R_y^\mathrm{gl}$ gates are applied globally to all atoms via RF-driven transitions between the two qubit states. The addressed $R_z$ operation is implemented by coupling one of the qubit states $\mid\uparrow\rangle$ to an auxiliary state $\vert a\rangle$ (e.g. another hyperfine state) using an off-resonant RF or microwave (MW) pulse. The AC Zeeman effect induces a phase shift to the qubit state, while the large detuning ensures that population remains in the qubit subspace~\cite{Wang2016}. The addressing laser locally shifts the auxiliary state of target atoms, leading to distinct $R_z$ rotations for target and non-target atoms.
This scheme offers two key advantages. First, spontaneous emission from the addressing laser is strongly suppressed due to its larger detuning from the qubit states. Second, the scheme provides striking insensitivity to addressing beam fluctuations to higher orders, making it applicable for small-spacing qubit arrays. 

\subsection{State-dependent single-atom detection}

Quantum gas microscopes have revolutionized ultracold atom experiments by enabling single-atom detection. Detection can be performed either in time-of-flight~\cite{Bucker2009, Bergschneider2018} or in situ~\cite{Bakr2009, Sherson2010, Omran2015, Xiang2025, Yao2025, Jongh2025}, with fidelities of single-atom detection exceeding 99\%. Distinct optical transitions allow individual spins to be resolved. From the acquired images, atomic momenta can be extracted from their positions in the waveguides, and atom-atom correlation functions can be directly reconstructed.

\subsection{Decoherence}

Decoherence in such a system arises from several sources. Inhomogeneous broadening leads to dephasing, which is largely suppressed by preparing atoms in the vibrational ground state at low temperature. Reversible dephasing is further suppressed using spin-echo sequences, yielding a coherence time $T_2'$ limited by magnetic-field fluctuations, laser intensity and phase noise, and atom loss. For the $^6$Li platform, the magnetic-field fluctuations dominate $T_2'$. To mitigate this effect we operate at high magnetic fields, where the chosen qubit states exhibit a reduced differential sensitivity of about 5~Hz/mG near the relevant Feshbach resonances. Given a field stability of $\sim $1~mG ($\sim 100~\mu$G), we expect a coherence time of $T_2' \sim 200~\text{ms}$ ($\sim 2~\text{s}$).

\section{Entanglement measurements}
\label{sec:Measurements}

In this section, we demonstrate the measurement of entanglement in spin, position–momentum, and path degrees of freedom, as well as hyperentanglement across multiple degrees of freedom. 

\subsection{Spin entanglement}

The experimental realization provides a direct matter-wave analog of Bohm’s version of the EPR Gedanken experiment, using two spin-$1/2$ fermions prepared in the spin-singlet state
\begin{equation}
|\Psi^-\rangle=\frac{1}{\sqrt{2}}\left(|\uparrow\rangle_L|\downarrow\rangle_R - |\downarrow\rangle_L|\uparrow\rangle_R\right)\,,
\end{equation}
which is a maximally entangled state. 

To verify the non-classical nature of this spin entanglement, a Bell test based on the Clauser–Horne–Shimony–Holt (CHSH) inequality can be performed. Spin measurements on each atom are carried out along directions specified by angles $\theta_L$, $\theta_{L'}$ for one atom, and $\theta_R$, $\theta_{R'}$ for the other. Experimentally, the measurement bases are defined by applying spin rotations about an axis in the equatorial plane of the Bloch sphere, prior to spin-resolved detection.

Quantum mechanics predicts that the correlation between outcomes is given by
\begin{equation}
\langle L_{\theta_L} R_{\theta_R} \rangle = -\cos(\theta_L - \theta_R)\,,
\end{equation}
where $L_{\theta_L}, R_{\theta_R} \in \{-1,+1\}$ are the binary spin outcomes projected along directions $\theta_L$ and $\theta_R$, respectively.

Choosing the standard CHSH angle configuration,
\begin{equation}
\theta_L = \frac{3\pi}{4},\quad \theta_R = \frac{\pi}{2},\quad \theta_{L'} = \frac{\pi}{4},\quad \theta_{R'} = 0\,,
\end{equation}
we obtain the following combination of spin correlation functions:
\begin{align}
S=&|\langle L_{3\pi/4} R_{\pi/2} \rangle - \langle L_{3\pi/4} R_0 \rangle| \notag \\
&+ |\langle L_{\pi/4} R_{\pi/2} \rangle + \langle L_{\pi/4} R_0 \rangle| = 2\sqrt{2}\,,
\label{eq:bell}
\end{align}
which exceeds the classical bound of 2 and thus violates the CHSH inequality.

Taking into account the fidelities of state preparation and detection, as well as additional reductions in visibility due to basis misalignment and dephasing, we predict an expected Bell violation of $S\approx2.45$, well above the classical bound of 2. This indicates that a clear Bell violation should be within reach under realistic conditions. Such a violation confirms the presence of genuine quantum entanglement between the spins of the two atoms. Importantly, this test closes the detection loophole due to the high fidelity, fine spatial resolution, and precise temporal control of spin-resolved detection at single-atom level. 

An alternative approach to verifying spin entanglement in the atom pair employs Wigner’s formulation of the Bell inequality~\cite{Wigner1970}, which reduces the number of required measurement settings but applies under slightly different assumptions. In either method, the experimental system offers a powerful platform for probing the fundamental nonlocal features of quantum mechanics with massive particles.

\subsection{Position-momentum entanglement}

In the spirit of the original Einstein-Podolsky-Rosen (EPR) protocol, momentum conservation enforces strong relative correlations between the two dissociated atoms. Although the individual positions and momenta of the dissociated atoms are uncertain, they ideally satisfy $x_1=x_2$ and $p_1=-p_2$, where $x_1$, $x_2$, $p_1$, and $p_2$ denote the positions and momenta of the two atoms. This continuous‑variable entanglement can be confirmed if the product of their conditional uncertainties falls below the Heisenberg limit:  
\begin{equation}
  \Delta(x_2\mid x_1)\,\Delta(p_2\mid p_1)
  < \hbar/2\,,
\end{equation}
where $\Delta(x_2\mid x_1)$ and $\Delta(p_2\mid p_1)$ denote the inferred standard deviations of $x_1-x_2$ and $p_1+p_2$, respectively. This scenario is closely analogous to the entangled photon pairs produced by SPDC~\cite{Howell2004}. 

Experimentally, immediately after dissociation and entry into the waveguides, we perform in-situ fluorescence imaging with a short pulse to measure the joint position distribution $P(x_1,x_2)$ and thereby determine $\Delta(x_2\mid x_1)$. In a separate but identically prepared ensemble, momentum measurements are implemented by allowing the atoms to undergo time‐of‐flight, after which we image their expanded spatial profiles to reconstruct the joint momentum distribution $P(p_1,p_2)$ and extract $\Delta(p_2\mid p_1)$. 
Considering a relative momentum spread of $\Delta(p_2\mid p_1) \sim 0.02\,\hbar k_{\rm rec}$ (with the recoil momentum 
$k_{\rm rec}=2\pi/671{\rm nm}$), and a typical imaging resolution of $\sigma_x \sim 1~\mu\text{m}$, we expect $\Delta(x_2\mid x_1)\,\Delta(p_2\mid p_1) \sim 0.2\,\hbar < \hbar/2$, which lies below the Heisenberg bound. 
A sufficient number of experimental repetitions will be conducted to achieve the statistical precision required to resolve these quantum‐mechanical signatures, including continuous-variable entanglement and squeezing.

\subsection{Path entanglement}

\begin{figure}[!tb]
\center
\includegraphics[width=1\columnwidth]{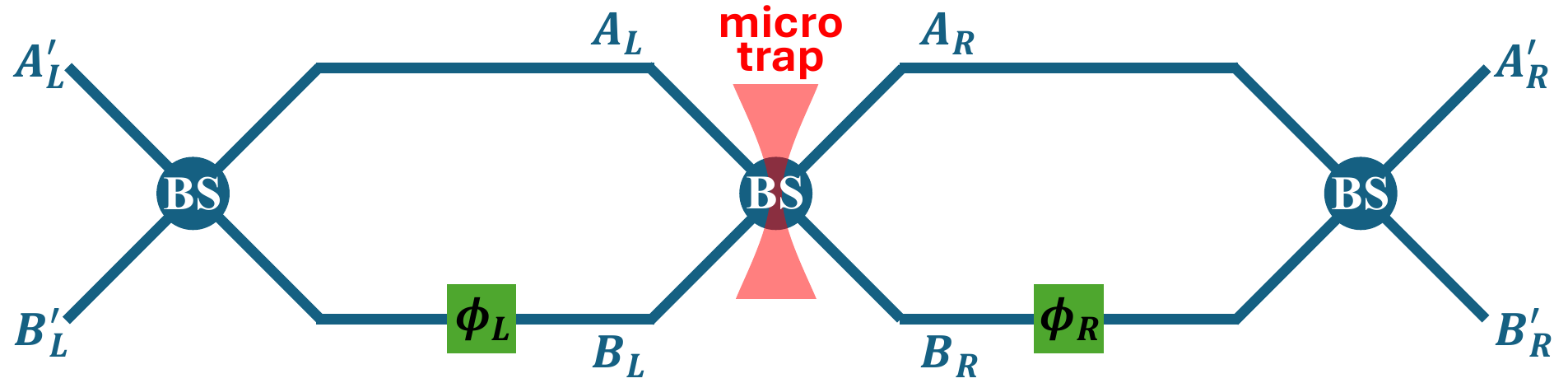}
\caption{Generation of maximally entangled path states from the dissociation of a single molecule and their verification via two-atom interferometry with tunable phase shifts. 
}
\label{fig:path}
\end{figure}

By replacing the optical lattice with programmable light modulators (SLMs or DMDs), one can create more flexible, customized waveguides and directly implement linear atom-optical elements within them. This protocol also enables the generation and manipulation of other forms of entanglement, such as path entanglement~\cite{Zhao2007}.

As illustrated in Fig.~\ref{fig:path}, two waveguides intersect at the center, coinciding with the microtrap that prepares a single diatomic molecule. Upon the dissociation of the molecule, the microtrap is turned off, and the center‐of‐mass wave packet of the two dissociated atoms enters two spatially separated “left” and “right“ regions. A 50:50 beam splitter (BS) then couples the atoms into paths $A_L/R,B_L/R$. By momentum conservation, the two‐atom state immediately after dissociation is a coherent superposition, 
\begin{equation}
\ket{\Psi_{\rm path}}=\frac{1}{\sqrt{2}}\bigl(\ket{A}_L\ket{B}_R - \ket{B}_L\ket{A}_R\bigr)\,,
\end{equation}
establishing maximal path entanglement across four spatial modes. This entangled state is insensitive to the precise temporal shape of the dissociation pulse. 

The path entanglement can be verified by observing a two-atom interferometer, formed by recombining $A_L,B_L$ (or $A_R,B_R$) via additional beam splitters near both ends. A controllable phase shift $\phi_L$ (or $\phi_R$) is applied between the two paths on each side. Joint detection probabilities at the outputs exhibit high‐visibility fringes given by  
\begin{equation}
P_{\pm}(\phi_L,\phi_R)
=\frac{1}{2}\Bigl[1 \pm \cos\phi_L\,\cos\phi_R\Bigr],
\end{equation}
where the signs indicate detector outcomes in each interferometer.  
In realistic implementations, the visibility of the path-entanglement fringes is degraded from the ideal unity due to a combination of experimental imperfections, including non-ideal state preparation and detection, background scattering, path imbalance, and finite precision in the realization of beam splitters and phase shifts. Taking these contributions into account, we simply estimate a net visibility of $V \approx 0.8$, which remains sufficient to demonstrate path entanglement within the proposed configuration.

Another scheme for generating both momentum and path entanglement in an atom pair is, during (at the end of) the molecular dissociation at the center of a single waveguide, to rapidly split the waveguide transversely into a double well of two parallel waveguides (paths $A$ and $B$)~\cite{Hohenester2007, Borselli2021, Kuriatnikov2025}. The two atoms move apart with opposite momenta along the split waveguide, the occupation of the transverse modes (paths) with predetermined correlations ~\cite{Borselli2021} reflecting the symmetry in the molecular dissociation. The resulting entanglement can be verified via two-particle interference. 

\subsection{Hyperentanglement}

By combining the internal (spin) and external (path) degrees of freedom of the atoms, a hyperentangled state can be prepared in an enlarged Hilbert space. Since spin and path can be manipulated independently, a single atom pair can encode four qubits, enabling the realization of a hyperentangled Greenberger–Horne–Zeilinger (GHZ) state. This is directly analogous to GHZ states demonstrated with photon pairs~\cite{Carvacho2017}. An idealized example of a four-qubit GHZ state in this system is: 
\begin{equation}
\ket{{\rm GHZ}} = \frac{1}{\sqrt{2}}\bigl(\ket{\uparrow_L A_L \downarrow_R B_R} - \ket{\downarrow_L B_L \uparrow_R A_R}\bigr)\,.
\end{equation}
Any projective measurement on one qubit collapses the global state and thereby determines the correlations among the remaining subsystems. By applying spin rotations and phase shifts between paths, other GHZ basis states can be obtained within the available four-qubit Hilbert space. Thus, without introducing additional particles, the hyperentangled encoding effectively realizes a GHZ entanglement structure within a single dissociated atom pair.

\section{Conclusions}
\label{sec:conclusion}

We have presented a deterministic entanglement source based on molecular dissociation, implementable with established neutral-atom techniques. This experimentally accessible platform enables atom-resolved characterization with near-unity detection fidelity, supports individual addressing with negligible crosstalk, and naturally scales through parallelization. Together, these capabilities provide a powerful foundation for future developments in matter wave based quantum technologies.

Looking ahead, deterministic entanglement sources of this type can serve as fundamental building blocks for integrated neutral-atom architectures. When combined with quantum operations in programmable atom-optics circuits and the robustness of large-scale AtomChip platforms~\cite{Folman2002, Reichel2011}, they open the door to a new generation of atomtronic devices and networks~\cite{Amico2021}. In such systems, near-field RF signals provide efficient initialization and precise control of individual atomic states, while reconfigurable optical potentials guide entangled matter waves through complex circuits. Reliable interconnections between EPR pairs, realized via controlled collisions or engineered junctions, would enable the assembly of larger entangled networks and the implementation of measurement-based quantum computation protocols.

By embedding deterministic entanglement sources into atomtronic architectures, one can envision compact matter-wave circuits that parallel photonic networks but exploit the unique advantages of massive particles: tunable interactions, long coherence times, and high-fidelity detection. Such integrated devices would not only provide a versatile platform for probing the foundations of quantum mechanics with unprecedented control, but also establish new pathways toward quantum technologies based on scalable networks of entangled neutral atoms.

\section*{Acknowledgements}

We thank Igor Mazets for valuable comments on entanglement measurement schemes, and Qi Liang and Pradyumna Paranjape for helpful discussions on experimental methods. 

This research was funded by the Austrian Science Fund (FWF), the ESPRIT grant ``Entangled Atom Pair Quantum Processor'' [grant DOI: 10.55776/ESP310]. This research was supported by the European Research Council: ERC-AdG ``Emergence in Quantum Physics'' (EmQ) under Grant Agreement No. 101097858, and Austrian Science Fund (FWF) stand alone ``Non-equilibrium dynamics in strongly interacting 1D quantum systems'' (NEqD-si1D) [grant DOI: 10.55776/P35390].  
For open access purposes, the author has applied a CC BY public copyright license to any author-accepted manuscript version arising from this submission.

\bibliography{Ref.bib}

\end{document}